\documentclass[aps,twocolumn,prb,showpacs,amsmath,amssymb,superscriptaddress]{revtex4}

\usepackage{graphicx}
\usepackage{dcolumn}
\usepackage{bm}
\usepackage[usenames]{color}

\begin{document}


\title{Are the renormalized band widths in 
TTF-TCNQ of structural or electronic origin? --- An angular 
dependent NEXAFS study.}

\author{M. Sing}
\email{michael.sing@physik.uni-wuerzburg.de}
\affiliation{Experimentelle Physik IV, Universit\"at W\"urzburg,
D-97074 W\"urzburg, Germany}
\author{J. Meyer}
\affiliation{Experimentelle Physik IV, Universit\"at W\"urzburg, 
D-97074 W\"urzburg, Germany}
\author{M. Hoinkis}
\affiliation{Experimentelle Physik IV, Universit\"at W\"urzburg, 
D-97074 W\"urzburg, Germany} \affiliation{Experimentalphysik II, 
Universit\"at Augsburg, D-86135 Augsburg, Germany}
\author{S. Glawion}
\affiliation{Experimentelle Physik IV, Universit\"at W\"urzburg, 
D-97074 W\"urzburg, Germany}
\author{P. Blaha}
\affiliation{Institute of Materials Chemistry, Vienna University
of Technology, A-1060 Vienna, Austria}
\author{G. Gavrila}
\affiliation{Institut f\"ur Physik, Technische Universit\"at 
Chemnitz, D-09107 Chemnitz, Germany}
\author{C.~S. Jacobsen}
\affiliation{Department of Physics, Technical University of Denmark, 
DK-2800 Lyngby, Denmark}
\author{R. Claessen}
\affiliation{Experimentelle Physik IV, Universit\"at W\"urzburg, 
D-97074 W\"urzburg, Germany}

\date{\today}

\begin{abstract}
We have performed angle-dependent near-edge x-ray absorption fine 
structure measurements in the Auger electron yield mode on the 
correlated quasi-one-dimensional organic conductor TTF-TCNQ in order 
to determine the orientation of the molecules in the topmost surface 
layer. We find that the tilt angles of the molecules with respect to 
the one-dimensional axis are essentially the same as in the bulk. 
Thus we can rule out surface relaxation as the origin of the 
renormalized band widths which were inferred from the analysis of 
photoemission data within the one-dimensional Hubbard model. Thereby 
recent theoretical results are corroborated which invoke long-range 
Coulomb repulsion as alternative explanation to understand the 
spectral dispersions of TTF-TCNQ quantitatively within an {\it 
extended} Hubbard model.
\end{abstract}

\maketitle


\section{Introduction}
One-dimensional (1D) electron systems represent a valuable test case 
for the understanding of unusual quantum many-body states. On 
theoretical grounds the enhanced correlations in 1D lead to a 
breakdown of the paradigmatic Fermi liquid picture involving new 
generic low-energy excitations. These manifest themselves in a 
dynamic decoupling of charge and spin degrees of freedom commonly 
referred to as spin-charge separation.\cite{Voit95a} While the 
low-energy physics is captured $e.~g.$ within the so-called 
Tomonaga-Luttinger model, which assumes linear band dispersions and 
therefore does not imply any intrinsic energy scale, the simplest 
model to explore generic 1D physics on the energy scale of the 
conduction band width is the 1D single-band Hubbard model 
(HM).\cite{Benthien04} The organic quasi-1D conductor TTF-TCNQ, 
where stacks of either the TCNQ or TTF planar molecules are 
effectively doped and become conducting by charge transfer from TTF 
to TCNQ, was the first material where angle-resolved photoemission 
spectroscopy (ARPES) was able to positively identify such 1D 
single-band HM spectral features as spinon and holon branches and 
the shadow band.\cite{Claessen02,Sing03b} However, in order to 
achieve a quantitative description of the experimental data a 
renormalization of the hopping integral $t$ at the surface by a 
factor of $\approx 2$ with respect to the bulk value from 
density-functional theory or estimates from bulk-sensitive 
measurements had to be assumed. At that time it was argued that due 
to the different Madelung potential at the surface the topmost 
molecular layer may be relaxed with tilting angles of the molecules 
against the 1D axis being different from the bulk. An alternative 
proposition has recently been elaborated on,\cite{Koch07} namely the 
necessity to incorporate the next-nearest neighbor Coulomb repulsion 
$V$ in a HM description in addition to the onsite Coulomb energy 
$U$.\cite{Bulut06b} This reduces the effective energy for two 
electrons to pass each other to $U-V$ and hence enhances the 
effective hopping integral $t_{eff}\approx tU/(U-V)$ without the 
need to assume a larger $t$. Indeed, it was shown by 
density-functional theory that $V\approx U/2$ within a 
stack.\cite{Koch07} However, these results are based on the bulk 
crystal structure and an experimental proof that no surface 
reconstruction is at work here is still lacking. There are few 
experimental techniques which in principle could give access to the 
molecular surface tilt such as quantitative low-energy electron 
diffraction (LEED) $IV$ measurements or scanning tunneling 
microscopy (STM). Nevertheless these techniques if applicable at all 
provide only indirect information and need theoretical simulations 
to interpret the experimental results. Here we employ a method 
widely used in surface physics to determine orientations of adsorbed 
molecules and apply it to the surface of TTF-TCNQ, namely angular 
dependent near-edge x-ray absorption fine structure (NEXAFS) 
spectroscopy.\cite{Stoehr96}

\section{Experimental and technical details}

The NEXAFS measurements were performed at the PM-3 beamline of BESSY  
(Berlin, Germany) using its MUSTANG endstation. The energy 
resolution amounted to 35\,meV at the N~$K$ edge and was even lower 
according to the characteristics of a plane-grating monochromator at 
the C~$K$- and S~$L$-edges. Clean sample surfaces were exposed 
through {\it in situ} cleavage of TTF-TCNQ single crystals at low 
temperatures. To reduce radiation damage all spectra were recorded 
at 60\,K, {\it i.~e.} above the charge density wave (CDW) phase 
transition temperature in the metallic phase. In order to achieve 
the required surface sensitivity the absorption signal was monitored 
in the Auger electron yield mode using a SPECS PHOIBOS~150 analyzer 
set to kinetic energies of 151\,eV, 268\,eV, and 381\,eV for S-, C-, 
and N-specific Auger lines, respectively. At these kinetic energies 
the electron mean free path is typically less than 
10\,{\AA}\cite{Stoehr96} which compares to the thickness of a single 
molecular layer in TTF-TCNQ of $c/2=9.23$\,{\AA} where $c$ is the 
lattice constant perpendicular to the surface (see below). Hence the 
absorption signal indeed stems essentially from the topmost 
molecular layer. All absorption spectra were corrected for the time 
and energy dependence of the incoming photon flux through division 
by the ring current and the drain current of a freshly sputtered Pt 
foil, respectively. The data were normalized in an energy region 
about 75\,eV above the absorption thresholds, where the final states 
are basically free-electron-like and hence isotropic. Great care was 
exercised to avoid sample degradation upon irradiation. The 
irradiation exposure time of one and the same sample was about 
40\,min before significant changes in the photoemission spectra of 
the S~2$p$ core level were detected which turned out to monitor most 
sensitively photon-induced damages.

In a molecular orbital picture NEXAFS involves the resonant x-ray 
excitation of a $K$- or $L$-shell electron into unoccupied low-lying 
molecular states governed by dipole selection rules. These project 
out the $s$-, $p$-, or $d$-like components of the antibonding 
$\sigma^*$ or $\pi^*$ final states at the excited atom which makes 
NEXAFS site- and element-specific. The idea to determine the 
molecular orientation of molecules on or at surfaces is based on the 
polarization dependence of the involved dipole matrix elements. 
Maximum absorption is obtained when the electric field vector $\bf 
E$ of the linearly polarized x-rays points along a vector-like 
orbital or lies within a plane-like 
orbital.\cite{footnote_vectororbitals} The absorption intensity then 
scales as $\cos^2 \delta$ where $\delta$ denotes the angle between 
$\bf E$ and the direction of a vector-like orbital or the plane of a 
plane-like orbital. Detailed expressions for the angular dependence 
of the $K$-shell absorption signal are given in 
Ref.~\onlinecite{Stoehr87} considering both the polar and azimuthal 
orientation of molecular orbitals with respect to the sample surface 
normal and in addition the degree of polarization. We note that in 
our case the expressions are simplified and become applicable also 
for $L$-shell absorption since the azimuthal orientation of the TTF 
and TCNQ molecules can be assumed to be fixed and the variation of 
the x-ray light incidence angle is restricted to the plane spanned 
by the sample surface normal and the 1D axis. If one moreover takes 
into account that the molecules from adjacent stacks of the same 
kind are tilted in opposite directions one finally arrives at the 
following expressions for the angular dependence of vector and plane 
like orbitals, respectively:
\begin{eqnarray}
I_v&=&C\cdot(\cos^2\theta\cdot\cos^2\alpha+\sin^2\theta\cdot\sin^2\alpha)\label{Eq1}\\
I_p&=&C^{\prime}\cdot(P\cdot(1-\cos^2\theta\cdot\cos^2\gamma-\nonumber\\
& &-\sin^2\theta\cdot\sin^2\gamma)+(1-P))\label{Eq2}
\end{eqnarray}
Here $\theta$ is the angle between the direction of the incident 
x-rays and the sample surface, $\alpha$ is the angle between the 
sample surface normal and the vector pointing in the direction of a 
vector-like orbital, $\gamma$ denotes the angle between the sample 
surface normal and the normal vector of a plane-like orbital, $P$ is 
the degree of polarization, and $C$ and $C^{\prime}$ are 
normalization constants.\cite{Stoehr87} Note that $\alpha$ and 
$\gamma$ are related to the corresponding tilt angles of the 
molecules through $90^{\circ}-\alpha$ and $90^{\circ}-\gamma$, 
respectively. The tilt angles can be determined by this method to an 
accuracy of about $\pm 10^{\circ}$.  

In order to study the effect of surface relaxation on the hopping 
integral $t$ band calculations were performed within the standard 
density-functional theory (DFT) using the generalized gradient 
approximation (GGA).\cite{Perdew96} We employed the self-consistent 
full-potential linearized augmented plane wave (LAPW) method as 
implemented in the WIEN2k code.\cite{Blaha01}

\section{Structural and electronic properties of TTF-TCNQ}
\begin{figure}
\includegraphics[width=8cm]{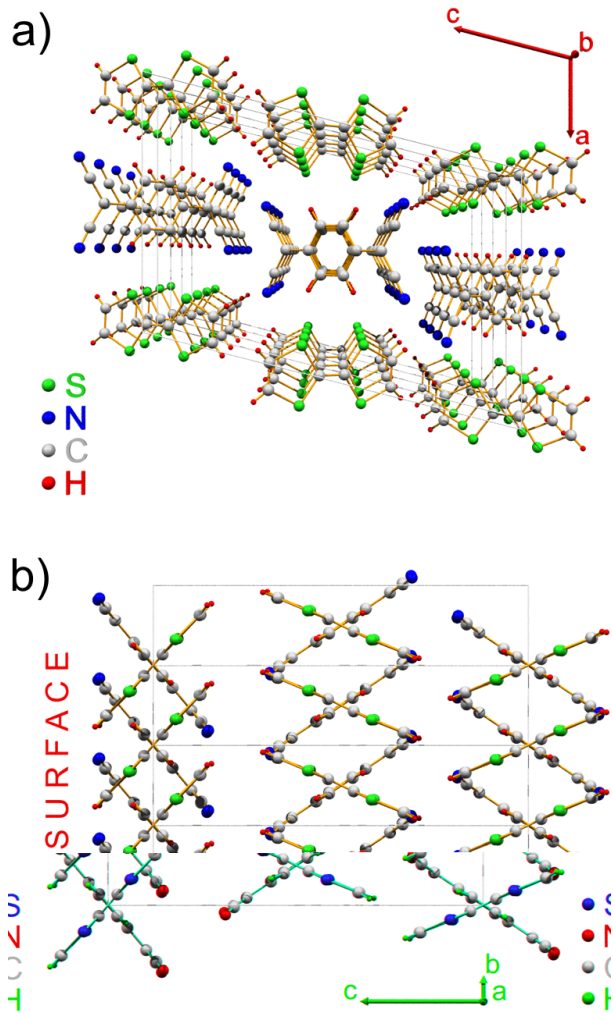}
\caption{\label{structure} (Color online) (a) Central projection 
view of the crystal structure of TTF-TCNQ along the $b$ axis. (b) 
Parallel projection view along $a$ with a hypothetical change in the 
molecule tilt angles at the surface. For details see text.}
\end{figure}

\begin{figure}
\includegraphics[width=8cm]{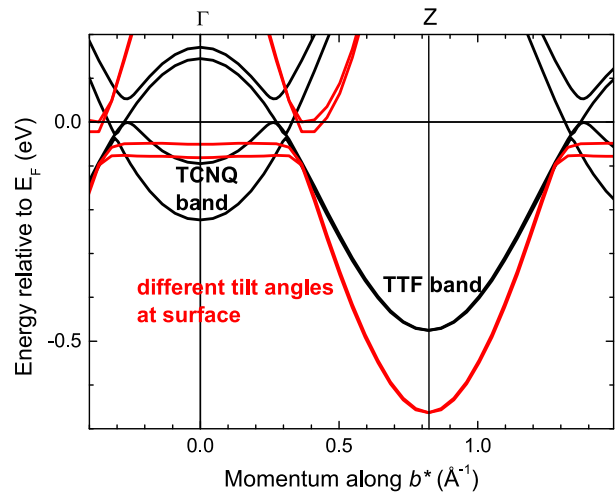}
\caption{\label{ldabands} (Color online) Bulk GGA band structure of 
TTF-TCNQ along the 1D direction for the real (black solid lines) and 
a hypothetical (red solid lines) band structure based upon enlarged 
molecule tilt angles. For details see text.}
\end{figure}

TTF-TCNQ 
(tetra\-thia\-fulva\-lene-tetra\-cyano\-quino\-di\-me\-thane) 
crystallizes in a monoclinic structure (see Fig.~\ref{structure}), 
space group $P2_1/c$, with lattice parameters $a=12.298$\,{\AA}, 
$b=3.819$\,{\AA}, $c=18.468$\,{\AA}, and 
$\beta=104.46^{\circ}$.\cite{Kistenmacher74} It consists of stacks 
of either TTF or TCNQ molecules which run along the crystallographic 
$b$ axis. Quasi-one-dimensional bands are formed through overlapping 
$\pi$ type molecular orbitals which extend over the entire 
molecules. 
Maximum hopping probability along the stacks and hence gain in 
kinetic energy is accomplished by a tilt of the essentially planar 
and rigid molecules around the $a$ axis by $24.5^{\circ}$ and 
$34.0^{\circ}$ for the TTF and TCNQ molecules, respectively. The 
tilt angle is oppositely oriented between adjacent stacks. 
Metallicity arises through charge transfer of about $0.6e$ per TTF 
to each TCNQ molecule. A detailed description of the strongly 
anisotropic band structure as obtained by GGA calculations can be 
found elsewhere.\cite{Sing03b} Here we focus on the TTF- and 
TCNQ-derived conduction bands just below the Fermi energy $E_F$ 
along $b$, i.~e. along $\Gamma$Z in the 1st Brillouin zone (black 
lines in Fig.~\ref{ldabands}). Around $\Gamma$ a TCNQ-derived band 
doublet (split by hybridization of the electronic states of the two 
stacks per unit cell) can be identified while the corresponding 
almost degenerate TTF-derived doublet is found towards the Brillouin 
zone boundary. What is important in our context is the fact that the 
experimental ARPES dispersions indicate a bandwidth larger by a 
factor of $\approx 2$ with respect to band 
theory.\cite{Claessen02,Sing03b} This discrepancy might be 
reconciled by the assumption of a relaxed topmost molecular layer. 
At the surface the Madelung potential and the polarization screening 
are different from the bulk which may allow for different molecule 
tilt angles to minimize energy. Larger tilt angles lead to a smaller 
intermolecular distance $d$ and hence an increased transfer integral 
which is counteracted by a reduced area overlap through a larger 
lateral offset $l$ of adjacent molecules in a stack. Assuming 
explicitly a $1/{d^3}$-dependence 
\cite{Andersen79,footnote_transferintegral} of the transfer integral 
this simple consideration leads to estimates for renormalized tilt 
angles as required to reproduce the experimental band widths, {\it 
viz.} $50^{\circ}$ and $48.3^{\circ}$ for TCNQ- and TTF-molecules, 
respectively. In order to explore this idea of a possible surface 
relaxation further we have performed new DFT calculations for a 
hypothetical bulk structure based on these tilt angles. While these 
indeed almost replicate the observed larger TTF-derived conduction 
band width the occupied part of the TCNQ band around $\Gamma$ comes 
out to have hardly any dispersion. Probably this is due to the 
complex nodal structure of the TCNQ-LUMO\cite{Koch07} perpendicular 
to the molecular plane which was neglected in the above qualitative 
consideration but can modulate the otherwise slowly varying 
functional behavior of $|t|$ in an oscillating manner causing one or 
several minima.\cite{Kazmaier94} In any case, one can induce from 
the above that (i) a molecular surface relaxation with $d$ and $l$ 
values in a realistic parameter regime can indeed be responsible for 
the enhanced experimental band widths and (ii) that a sizeable 
increase in the tilt angles by $\sim 20^{\circ}$ would be necessary. 
We add that it was briefly noted in the context of recent GGA 
calculations that a structural optimization for the outer molecular 
layer keeping the others fixed found changes in the molecule 
inclination angles though with minor consequences for the electronic 
dispersions along $b$.\cite{Fraxedas03}

\section{Angular-dependent NEXAFS spectra}
The site and element sensitivity of NEXAFS enables the determination 
of the TTF and TCNQ surface tilt angles separately recording the 
S~$L$ and N~$K$ absorption edges, respectively. On the other hand 
the angular dependence of the C~$K$ edge spectra contains 
information about the orientation of both molecules provided that 
the numerous resonances can be disentangled. In order to accurately 
extract the angular dependence of a certain resonance feature the 
NEXAFS spectra have to be analyzed using a suitable fitting scheme. 
We followed the approach of Outka and St{\"o}hr \cite{Outka88} and 
fitted the spectra after a proper background subtraction with a 
minimal number of Gaussian- and Lorentzian-like peaks. Absorption 
into the free-electron continuum was accounted for by an error 
function step where appropriate, whose position and width reflects 
the binding energy and the relative chemical shift\cite{Sing03} due 
to the various inequivalent atoms in the structure. No "hidden" 
peaks were included in the fitting at a certain angle unless there 
was a definitive indication for them in the spectra at neighboring 
angles and a remnant still seen in the spectrum to be fitted. 
\begin{figure}
\includegraphics[width=8cm]{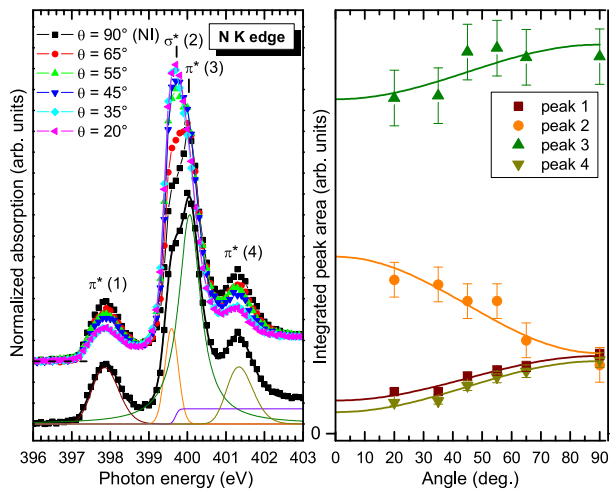}
\caption{\label{NKedge} (Color online) Left panel: N~$K$ edge NEXAFS 
spectra of TTF-TCNQ for various light incidence angles $\theta$. In 
the lower part the normal incidence (NI) spectrum is reproduced 
together with the corresponding fit analysis from which the integral 
peak intensities are obtained. Right panel: Integrated peak areas 
for the peaks as in the fit of the left panel as a function of light 
incidence angle (dotted curve). The solid lines mark fits according 
to the model as described in the text.}
\end{figure}
Ideally, during the fitting of a complete angle series the 
parameters for all peaks (including the continuum step) are  
determined once and for all for a particular spectrum and then held 
fixed except for the ones describing the relative integral weight of 
each component. In practice, this cannot be sustained all over, 
especially for the C~$K$ edge where 9 crystallographically different 
C~atoms contribute and a unique identification of each resonance for 
all angles becomes difficult. Moreover, a small relaxation (within a 
few percent) of all parameters was permitted in order to achieve 
better fits. For peak assignment we relied on the comprehensive 
NEXAFS study of Fraxedas {\it et al.} on highly textured TTF-TCNQ 
films.\cite{Fraxedas03} Note that throughout this paper $\sigma^*$ 
and $\pi^*$ resonances correspond to plane- and vector-like 
molecular orbitals within the classification scheme of 
Ref.~\onlinecite{Stoehr87}. Employing Eq.~\ref{Eq2} to fit the 
angular intensity variation of a resonance excitation into a 
plane-like orbital we assumed a polarization degree $P\approx 1$

In the left panel of Fig.~\ref{NKedge} N~$K$ absorption spectra are 
displayed for various light incidence angles $\theta$. On unbiased 
inspection four resonances can be discerned which are numbered and 
denoted according to their symmetry. Their intensity variations with 
different light incidence angles are clearly seen. The N~$K$ edge 
spectra represent the "clean" case, $i.~e.$ all spectra can be 
fitted with one single set of underlying components. An example of 
such a fit analysis with the total fit curve and its decomposition 
into the various constituents is depicted in the lower part of the 
panel for $\theta = 90^{\circ}$. The integral intensities of all 
four peaks are plotted in the right panel of Fig.~\ref{NKedge} as a 
function of $\theta$ together with regression curves according to 
Eqs.~\ref{Eq1} and \ref{Eq2}. As tilt angles for the TCNQ molecules 
one thus obtains $33.2^{\circ}$ ($34.0^{\circ}$, $42.8^{\circ}$, 
$28.6^{\circ}$) for peak~1 (2, 3, 4). These values are in fair 
agreement with the number $34.0^{\circ}$ from the bulk. The somewhat 
higher value derived from the intensity variation of peak~3 probably 
reflects the insufficient approximation of the underlying spectral 
weight by just one peak as can be seen from its relative broad 
Lorentzian shape. Actually, most of the resonances which are denoted 
here comprise more than one molecular orbital with the same 
symmetries.\cite{Fraxedas03} However, this shortcoming has to be 
accepted following the directive to use a minimal set of peaks, for 
all of which clear indication is to be seen in at least one 
spectrum.

\begin{figure}
\includegraphics[width=8cm]{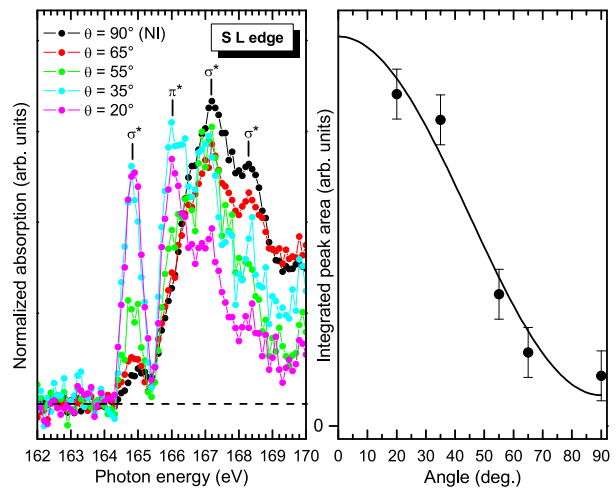}
\caption{\label{SLedge} (Color online) Left panel: S~$L$ edge NEXAFS 
spectra of TTF-TCNQ for various light incidence angles $\theta$. 
Right panel: Integrated area for the pronounced $\sigma^{*}$ 
resonance peak at 164.8\,eV (dotted curve). The solid line depicts a 
fit according to the model as described in the text.}
\end{figure}
Figure~\ref{SLedge} shows S~$L$ edge NEXAFS spectra which are 
specific for TTF. As before distinct intensity variations as a 
function of $\theta$ are observed for at least four resonances which 
again are marked according to their symmetry in the plot. A detailed 
fit analysis is hampered here for two reasons. First the 
signal-to-noise ratio is much worse in the Auger yield detection 
mode as compared to the N~$K$ edge which however cannot be easily 
compensated by longer measuring times due to the comparatively rapid 
radiation induced sample degradation. Second the overall spectral 
shape is much broader and the individual resonance peaks are much 
less well separated than in the case of the N~$K$ edge spectra 
except for the $\sigma^*$ resonance at 164.8\,eV. Therefore we only 
fitted this single peak for the evaluation of the TTF tilt angle. 
From the fit of the angle-dependent integral intensity which is 
displayed in the right panel of Fig.~\ref{SLedge} a value of 
$15.7^{\circ}$ is obtained again in reasonable agreement with the 
bulk value of $24.5^{\circ}$ and at variance with the expectation in 
case of a sizeable surface relaxation.

\begin{figure}
\includegraphics[width=8cm]{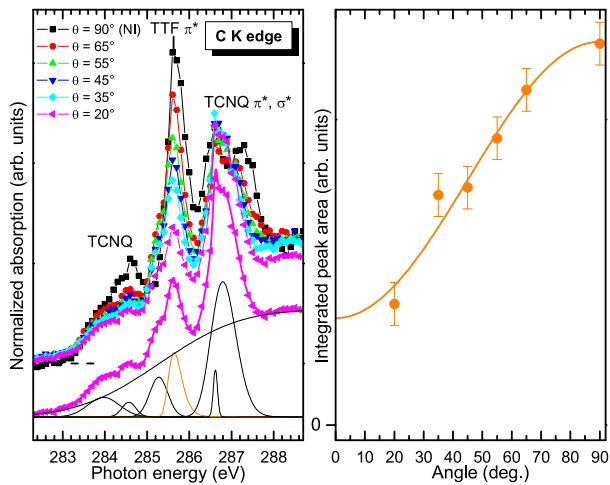}
\caption{\label{CKedge} (Color online) Left panel: C~$K$ edge NEXAFS 
spectra of TTF-TCNQ for various light incidence angles $\theta$. In 
the lower part the spectrum for $\theta=20^{\circ}$ is reproduced 
together with the corresponding fit analysis from which the integral 
peak intensities were obtained. Right panel: Integrated peak area 
for the peak marked in orange in the left panel as a function of 
light incidence angle (dotted curve). The solid line depicts a fit 
according to the model as described in the text.}
\end{figure}
The results obtained so far for the TTF and TCNQ tilt angles might 
be corroborated from an analysis of the C~$K$ spectra if one 
succeeds to trace the intrinsic angular dependence of a resonance 
located either on the TTF or TCNQ molecules. In Fig.~\ref{CKedge} 
C~$K$ spectra taken at different light incidence angles are plotted. 
By comparison of C~$K$ edge spectra for TTF, TCNQ, and TTF-TCNQ 
shown in Ref.~\onlinecite{Fraxedas03} we can grossly label few 
structures in our data according to their symmetry and origin from 
either TTF or TCNQ. From the fit analysis --- a representative 
regression curve is plotted in the lower part of the left panel in 
Fig.~\ref{CKedge} 
--- it turns out that a unique tracking is only possible for the 
most pronounced resonance structure at 285.6\,eV which is identified 
as a $\pi^*$ resonance on TTF. In the right panel the angular 
variation of its integral intensity is shown together with the 
corresponding fit curve giving a tilt angle of $27.8^{\circ}$. This 
again reflects the bulk value of $24.5^\circ$ in agreement with the 
analysis of the S~$L$ edge spectra instead of a strongly increased 
tilt for a relaxed surface.

\section{Discussion}
In the search of experimental realizations of generically 1D systems 
many claims have been made on the observation of certain {\it 
isolated} aspects such as spin-charge separation or the 
non-universal power-law suppression of spectral weight at the 
chemical potential. However, it is fair to say that a {\it 
comprehensive} theoretical and experimental picture where a 
sufficient number of pieces are put together thus exhibiting clear 
1D phenomenology has not yet been achieved for any of these systems. 
Probably for TTF-TCNQ such an understanding is farthest developed. 
There not only the electronic ARPES dispersions of the TCNQ related 
spectral weight could be quantitatively fitted by the spinon and 
holon branches of the 1D HM using the Bethe Ansatz 
solution\cite{Sing03b} and the pseudofermion dynamical 
theory\cite{Carmelo04} but also the overall spectral weight 
distribution could be reproduced by density-matrix 
renormalization-group (DDMRG) calculations.\cite{Benthien04} The 
underlying parameter set in both cases, {\it viz.} $U=1.96$\,eV and 
$t=0.4$\,eV, however, assumes a hopping integral $t$ which is 
enhanced by a factor of $\approx 2$ with respect to the DFT 
calculations. To justify this assumption it was argued from an 
experimental point of view that the topmost surface layer of 
TTF-TCNQ might be relaxed.\cite{Claessen02,Sing03b} On the other 
hand, further theoretical work has shown that the also observed 
transfer of spectral weight at $k_F$ over the entire conduction band 
width with increasing temperatures cannot be reconciled within this 
parameter set.\cite{Abendschein06,Bulut06b} The effective magnetic 
exchange $J_{eff}$ which defines a characteristic temperature for 
the onset of such an spectral-weight transfer would come out much 
too large within the 1D HM. This questions the assumption of a 
surface relaxation and hence an enhanced value of $t$ to explain the 
experimental data. Results by exact diagonalization\cite{Bulut06b} 
for $T=0$ found that the inclusion of the nearest-neighbor Coulomb 
repulsion $V$ in the extended Hubbard model offers a possible 
loophole since it effectively increases the bandwidths of the spinon 
and holon excitations to match the experimental dispersions while at 
the same time allowing for a smaller $t$ and hence a smaller 
$J_{eff}$ in better agreement with experiment. Indeed, Koch {\it et 
al.}\cite{Koch07} recently have reported on density-functional 
results which find that the long-ranged Coulomb repulsion even to 
the third nearest neighboring molecule within a stack is still about 
20\,\% of $U$. However, these results are based on the bulk crystal 
structure of TTF-TCNQ. In this context the NEXAFS work presented 
here provides the missing link which is needed to reconcile ARPES 
dispersions and $T$ dependence with theory, {\it viz.} HM and DFT 
calculations. Our results indicate that no surface relaxation, {\it 
i.~e.} no extra molecule tilt, occurs for either the TTF or the TCNQ 
stacks leaving the hopping integral $t$ at the surface essentially 
unaltered with respect to the bulk. These findings validate the work 
by Koch {\it et al.} which in turn corroborates the theoretical 
conclusion concerning the temperature-dependent redistribution of 
spectral weight above $J_{eff}$. It should not be concealed here 
that a theoretical description of the complete ARPES data is not yet 
accomplished. In particular, the TTF derived part of the data calls 
for a consistent treatment on the same footing. This should be able 
to explain {\it e.~g.} the seeming absence of the spinon-holon 
splitting although the electronic states of the TTF stacks should be 
more strongly correlated than those of the TCNQ stacks.
               
\section{Conclusions}
From the analysis of the angular dependence of the N~$K$, S~$L$, and 
C~$K$ NEXAFS spectra we arrive at a consistent picture where under 
the restrictions imposed by experiment the topmost molecular layer 
of TTF-TCNQ single crystals appears to be structurally unaltered 
with respect to the bulk. This result lends additional significance 
to the approach to explain the deviations between experimental and 
DFT band dispersions purely electronically. A refinement of the 
single-band 1D Hubbard model by inclusion of a next-neighbor Coulomb 
repulsion $V$ as proposed to explain the temperature dependence of 
the photoemission data thus seems plausible from an experimental 
point of view.

\begin{acknowledgments}
We are grateful to M. Sperling for technical assistance during the 
experiments at BESSY and thank A. Sch{\"o}ll for fruitful 
discussions. The experimental end station was funded under MUSTANG 
BMBF 05 KS4OC1/3.
\end{acknowledgments}


\begin{thebibliography}{20}
\bibitem[{foo({\natexlab{b}})}]{footnote_transferintegral}
\bibinfo{note}{Depending on the theoretical approach one finds different
  scaling behaviors for the transfer integral $t$ in the literature ranging
  from a power law decay as $1/d^2$ to an exponential decay. However, these
  discrepancies do not affect our qualitative arguments.}
\end{thebibliography}

\end{document}